\documentclass[a4paper]{article}

\usepackage[T1]{fontenc} 
\usepackage{float}       
\usepackage{helvet}      
\usepackage{mathptmx}    
\usepackage{rsc}         
\usepackage{setspace}    
\usepackage{graphics}
\usepackage{epsfig}
\AtBeginDocument{\doublespacing}

\newfloat{chart}{htbp}{loc}
\newfloat{graph}{htbp}{loh}
\newfloat{scheme}{htbp}{los}

\usepackage[version=3]{mhchem} 

\author{%
Olivier Mousis\thanks{%
Universit{\'e} de Franche-Comt{\'e}, Institut UTINAM, CNRS/INSU, UMR 6213, France; Universit\'e europ\'{e}enne de Bretagne; Universit{\'e} de Rennes 1, Institut de Physique de Rennes, CNRS, UMR 6251, France; 
E-mail: \texttt{olivier.mousis@obs-besancon.fr}
}%
\and
Jonathan I. Lunine\thanks{%
Dipartimento di Fisica, Universit{\`a} degli Studi di Roma ``Tor Vergata'', Rome, Italy%
}%
\and
Sylvain Picaud\thanks{%
Universit{\'e} de Franche-Comt{\'e}, Institut UTINAM, CNRS/INSU, UMR 6213, France%
}%
\and
Daniel Cordier\thanks{%
Ecole Nationale Sup{\'e}rieure de Chimie de Rennes, CNRS, UMR 6226, France; Universit\'e europ\'{e}enne de Bretagne; Universit{\'e} de Rennes 1, Institut de Physique de Rennes, CNRS, UMR 6251, France%
}%
}

\title{Volatile inventories in clathrate hydrates formed in the primordial nebula}

\begin{document}

\maketitle

\begin{abstract}

Examination of ambient thermodynamic conditions suggest that clathrate hydrates could exist in the martian permafrost, on the surface and in the interior of Titan, as well as in other icy satellites. Clathrate hydrates probably formed in a significant fraction of planetesimals in the solar system. Thus, these crystalline solids may have been accreted in comets, in the forming giant planets and in their surrounding satellite systems. In this work, we use a statistical thermodynamic model to investigate the composition of clathrate hydrates that may have formed in the primordial nebula. In our approach, we consider the formation sequence of the different ices occurring during the cooling of the nebula, a reasonable idealization of the process by which volatiles are trapped in planetesimals. We then determine the fractional occupancies of guests in each clathrate hydrate formed at given temperature. The major ingredient of our model is the description of the guest-clathrate hydrate interaction by a spherically averaged Kihara potential with a nominal set of parameters, most of which being fitted on experimental equilibrium data. Our model allows us to find that Kr, Ar and N$_2$ can be efficiently encaged in clathrate hydrates formed at temperatures higher than $\sim$ 48.5 K in the primitive nebula, instead of forming pure condensates below 30 K. However, we find at the same time that the determination of the relative abundances of guest species incorporated in these clathrate hydrates strongly depends on the choice of the parameters of the Kihara potential and also on the adopted size of cages. Indeed, testing different potential parameters, we have noted that even minor dispersions between the different existing sets can lead to non-negligible variations in the determination of the volatiles trapped in clathrate hydrates formed in the primordial nebula. However, these variations are not found to be strong enough to reverse the relative abundances between the different volatiles in the clathrate hydrates themselves. On the other hand, if contraction or expansion of the cages due to temperature variations are imposed in our model, the Ar and Kr mole fractions can be modified up to several orders of magnitude in clathrate hydrates. Moreover, mole fractions of other molecules such as N$_2$ or CO are also subject to strong changes with the variation of the size of the cages. Our results may affect the predictions of the composition of the planetesimals formed in the outer solar system. In particular, the volatile abundances calculated in the giant planets atmospheres should be altered because these quantities are proportional to the mass of accreted and vaporized icy planetesimals. For similar reasons, the estimates of the volatile budgets accreted by icy satellites and comets may also be altered by our calculations. For instance, under some conditions, our calculations predict that the abundance of argon in the atmosphere of Titan should be higher than the value measured by Huygens. Moreover, the Ar abundance in comets could be higher than the value predicted by models invoking the incorporation of volatiles in the form of clathrate hydrates in these bodies.

\end{abstract}

\section{Introduction}

Clathrate hydrates (hereafter clathrates) were discovered in 1810 by Sir Humphrey Davy. Initially considered as laboratory curiosities, it is only from the 1930s that the study of their formation conditions became of significant interest because of the clogging pipelines during transportation of gas under cold conditions. Clathrates are crystalline solids which look like ice and form when water molecules constitute a cage-like structure around small ``guest molecules''. The most common guest molecules in terrestrial clathrates are methane (the most abundant), ethane, propane, butane, nitrogen, carbon dioxide and hydrogen sulfide. Water crystallizes in the cubic system in clathrates, rather than in the hexagonal structure of normal ice. Several different clathrate structures are known, the two most common ones being named ``structure I'' and ``structure II''. In structure I, the unit cell is formed of 46 water molecules and can incorporate up to 8 guest molecules. In structure II, the unit cell consists  of 136 water molecules and can incorporate at most 24 guest molecules. 

The thermodynamic conditions prevailing in many bodies of the solar system suggest that clathrates could also exist in the martian permafrost\cite{Chastain07,Thomas09,Swindle09}, on the surface and in the interior of Titan as well as in other icy satellites\cite{Lunine1985,Hand06,Tobie06,Thomas07,Mousis08}. Moreover, it has been suggested that the activity observed in some cometary nuclei results from the dissociation of these crystalline structures\cite{Marboeuf10}. Generally speaking, clathrates probably participated in the formation of planetesimals in the solar system. Indeed, formation scenarios of the protoplanetary nebula invoke two main reservoirs of ices that took part in the production of icy planetesimals. The first reservoir, located within 30 Astronomical Units (AU) of the Sun, contains ices (mostly water ice) originating from the Interstellar Medium (ISM) which, due to their proximity to the Sun, were initially vaporized\cite{Chick97}. With time, the decrease of temperature and pressure conditions allowed the water in this reservoir to condense at $\sim$ 150 K (for a total gas pressure of $\sim$ 10$^{-7}$ bar) in the form of microscopic crystalline ice\cite{Kouchi94}. It is then considered that a substantial fraction of the volatile species were trapped as clathrates as long as free water ice was available within $\sim$ 30 AU in the outer solar nebula\cite{Mousis00}. On the other hand, remaining volatiles that have not been enclathrated due to the lack of available water ice have probably formed pure condensates at lower temperatures in this part of the nebula\cite{Mousis09a,Mousis09b}. The other reservoir, located at larger heliocentric distances, is composed of ices originating from ISM that did not vaporize when entering into the disk. In this reservoir, water ice was essentially in the amorphous form and the other volatiles remained trapped in the amorphous matrix\cite{Owen99,Notesco05}. Consequently, icy planetesimals formed at heliocentric distances below 30 AU mainly agglomerated from clathrates while, in contrast, those produced at higher heliocentric distances (i.e. in the cold outer part of the solar nebula) are expected to be formed from primordial amorphous ice originating from ISM. Thus, clathrates may have been accreted in comets, in the forming giant planets and in their surrounding satellite systems\cite{Lunine1985,Mousis00,Gautier01a,Gautier01b,Alibert05,Alibert07,Mousis04,MG04,MA06,Mousis06,Mousis09a,Mousis09b}.

During the twentieth century, many theoretical and experimental studies allowed characterization of the crystalline structures of the most common clathrates. Meanwhile, a classification has been established to identify the nature of the clathrate and the form of occupation of the trapped molecules (single clathrate, multiple guest clathrates, etc). From the knowledge of the structure of clathrates, predictive rigorous methods have been developed to determine their thermodynamic properties. In particular, van der Waals \& Platteeuw\cite{vwp59} laid the foundations of a statistical thermodynamics model to determine the properties of clathrates. This method is an excellent modern example of the use of statistical thermodynamics to predict macroscopic quantities such as temperature and pressure, using the microscopic properties like potential interactions. This approach, used today in the industry and in science, has saved substantial experimental efforts for the determination of  i) the equilibrium pressure of a clathrate formed from various mixtures and ii) the mole fraction of the different species trapped in the clathrate from a given fluid phase.

In the present work, we use a statistical thermodynamic model derived from the approach of van der Waals \& Platteeuw in order to investigate the composition of clathrates that may have formed in the solar nebula. Indeed, many works  published in the last decade and detailing the formation conditions of ices in the solar nebula have neglected the possibility of multiple guest trapping in clathrates\cite{Gautier01a,Gautier01b,Hersant04,Hersant08,Mousis06,Alibert07,Mousis09b}. In our approach, we consider the formation sequence of the different ices occurring during the cooling of the solar nebula and that is usually used to describe the process by which volatiles are trapped in planetesimals\cite{Gautier01a,Gautier01b,Hersant04,Hersant08,Mousis06,Alibert07,Mousis09b} (see Fig. \ref{cool}). We then determine the fractional occupancies of guests in each clathrate formed at given temperature. Similarly to papers following the approach of van der Waals \& Platteeuw, our model is based on the use of intermolecular potentials, which themselves depend on parameters describing the interaction between the molecule and the cage, called ``Kihara parameters''. Because the models are extremely sensitive to the choice of these parameters, and because different sets of data exist in the literature\cite{PP72a,PP72b,DP82,Sloan98,Jager2001,Kang01,Sloan08}, we examine the influence of the interaction potential parameters on our calculations. Moreover, it has been shown that the size of the clathrate cages depends on their formation temperature\cite{Shpakov97,Belosludov03,Takeya06}. This also leads us to investigate the influence of the size of cages on the resulting composition of clathrates formed in the low pressure and temperature conditions of the nebula.

The paper is organized as follows. In Section \ref{seq}, we briefly depict the formation sequence of the different ices, including clathrates, in the outer solar nebula. In Section \ref{model}, we describe the statistical model based on the work of van der Waals and Platteeuw, and used to calculate the precise composition of clathrates. In Section \ref{predict}, we define an optimal set of Kihara parameters used in our statistical model in order to calculate the nominal fractional occupancies of the different guests incorporated in clathrates formed in the solar nebula. Section \ref{sens} is devoted to the investigation of the influence of the potential parameters and structural characteristics (i.e. size of cages) of clathrates on our calculations of the fractional occupancies. In Section \ref{imp}, we discuss the implications of our calculations for the composition of the outer solar system. In particular, we focus our attention on the volatile abundances in the atmosphere of Saturn, the issue of the measured argon deficiency in Titan and the prediction of the noble gas content in comets. Section \ref{sum} is devoted to the summary and discussion of our results.

\section{Formation of clathrates in the primordial nebula}
\label{seq}

In the present work, we focus our attention on the formation sequence of the different ices produced in the giant planets formation zone in the primordial nebula. Once formed, these ices will add to the composition of the planetesimals accreted by the giant planets and their surrounding satellites during their growth. The composition of the initial gas phase of the disk is defined as follows: we assume that the abundances of all elements, including oxygen, are protosolar\cite{Lodders03} and that O, C, and N exist only under the form of H$_2$O, CO, CO$_2$, CH$_3$OH, CH$_4$, N$_2$, and NH$_3$. The abundances of CO, CO$_2$,  CH$_3$OH, CH$_4$, N$_2$ and NH$_3$ are then determined from the adopted CO/CO$_2$/CH$_3$OH/CH$_4$ and N$_2$/NH$_3$ gas phase molecular ratios. Once the abundances of these molecules are fixed, the remaining O gives the abundance of H$_2$O. Concerning the distribution of elements in the main volatile molecules, we set CO/CO$_2$/CH$_3$OH/CH$_4$ = 70/10/2/1 in the gas phase of the disk, values that are consistent with the ISM measurements considering the contributions of both gas and solid phases in the lines of sight\cite{Frerking82,Ohishi92,Ehrenfreund00,Gibb00}. In addition, S is assumed to exist in the form of H$_2$S, with H$_2$S/H$_2$ = 0.5 $\times$ (S/H$_2$)$_{\odot}$\footnote{(S/H$_2$)$_{\odot}$ means the solar abundance of S relative to H$_2$. $\odot$ is the astronomical symbol for the Sun.}, and other refractory sulfide components\cite{Pasek05}. We also consider N$_2$/NH$_3$ = 10/1 in the nebula gas-phase, a value compatible with thermochemical models of the solar nebula\cite{Lewis80} and with observations of cometary comae\cite{Hersant08}. In the following, we adopt these mixing ratios as our nominal model of the solar nebula gas phase composition (see Table \ref{lodders}).

\begin{table}
\caption[]{Gas phase abundances in the solar nebula.}
\begin{center}
\begin{tabular}{lclc}
\hline
\noalign{\smallskip}
Species X &  (X/H$_2$)  & species X  & (X/H$_2$) \\	
\noalign{\smallskip}
\hline
\noalign{\smallskip}
O		& $1.16 \times 10^{-3}$		& N$_2$		&  $7.62 \times 10^{-5}$ \\
C		& $5.82 \times 10^{-4}$ 		& NH$_3$   	&  $7.62 \times 10^{-6}$ \\
N   		& $1.60 \times 10^{-4}$    		& CO      		& $4.91 \times 10^{-4}$ \\
S        	& $3.66 \times 10^{-5}$		& CO$_2$  	& $7.01 \times 10^{-5}$ \\
Ar       	& $8.43 \times 10^{-6}$		& CH$_3$OH  	& $1.40 \times 10^{-5}$ \\ 	
Kr       	& $4.54 \times 10^{-9}$		& CH$_4$  	& $7.01 \times 10^{-6}$ \\ 
Xe       	& $4.44 \times 10^{-10}$		& H$_2$S    	& $1.83 \times 10^{-5}$ \\
H$_2$O  	&  $5.15 \times 10^{-4}$		&							     \\
\hline
\end{tabular}
\end{center}
Elemental abundances derive from Lodders (2003). Molecular abundances result from the distribution of elements following the approach given in the text.
\label{lodders}
\end{table}

\begin{figure}
\centering
\resizebox{\hsize}{!}{\includegraphics[angle=0]{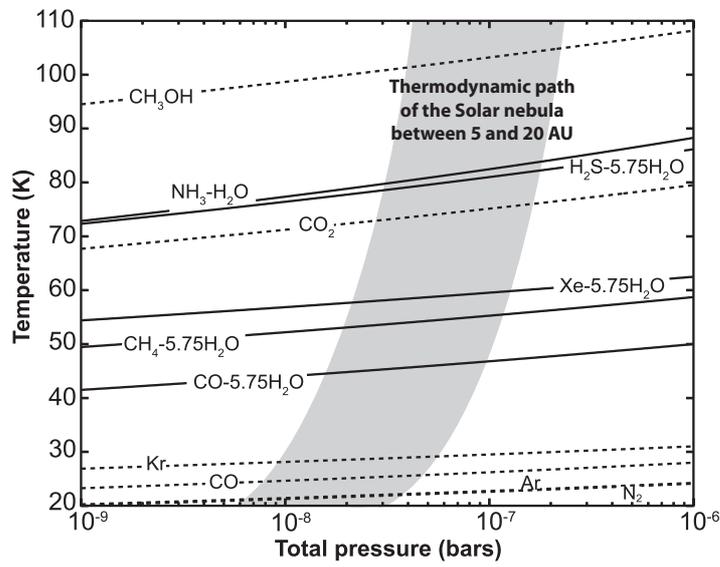}}
\caption{Equilibrium curves of NH$_3$-H$_2$O hydrate, H$_2$S, Xe, CH$_4$ and CO clathrates (solid lines), CH$_3$OH, CO$_2$, Kr, CO, Ar and N$_2$ pure condensates (dotted lines), and thermodynamic path followed by the solar nebula between 5 and 20 AU as a function of time, respectively, assuming a full efficiency of clathration. Abundances of various elements are solar, with CO/CO$_2$/CH$_3$OH/CH$_4$ = 70/10/2/1, H$_2$S/H$_2$ = 0.5 $\times$ (S/H$_2$)$_{\odot}$, and N$_2$/NH$_3$ = 10/1 in the gas phase of the disk. Species remain in the gas phase above the stability curves. Below, they are trapped as clathrates or simply condense.}
\label{cool}
\end{figure}

The process by which volatiles are trapped in icy planetesimals, illustrated in Fig. \ref{cool}, is calculated using the stability curves of stochiometric hydrates, clathrates and pure condensates, and the thermodynamic path (hereafter cooling curve) detailing the evolution of temperature and pressure between 5 and 20 AU roughly corresponding to the formation locations of the giant planets in the solar nebula \footnote{Recent models of giant planets formation show that instead of forming at 20 and 30 AU, i.e. their current orbits, Uranus and Neptune underwent most of their growth among proto-Jupiter and proto-Saturn and were scattered outward when Jupiter acquired its massive gas envelope, and subsequently evolved toward their present orbits\cite{Thommes02,Tsiganis05,Lykawka09}.}. We refer the reader to the works of Papaloizou \& Terquem\cite{Papaloizou99} and Alibert et al.\cite{Alibert05b} for a full description of the turbulent model of accretion disk used here. The stability curves of hydrates and clathrates derive from Lunine \& Stevenson's compilation\cite{Lunine1985} of published experimental work, in which data are available at relatively low temperatures and pressures. On the other hand, the stability curves of pure condensates used in our calculations derive from the compilation of laboratory data given in the CRC Handbook of Chemistry and Physics\cite{Lide02}. The cooling curve intercepts the stability curves of the different ices at particular temperatures and pressures. For each ice considered, the domain of stability is the region located below its corresponding stability curve. The clathration process stops when no more crystalline water ice is available to trap the volatile species. Note that, in the pressure conditions of the solar nebula, CO$_2$ is the only species that crystallizes at a higher temperature than its associated clathrate. We then assume that solid CO$_2$ is the only existing condensed form of CO$_2$ in this environment. In addition, we have considered only the formation of pure ice of CH$_3$OH in our calculations since, to the best of our knowledge, no experimental data concerning the stability curve of its associated clathrate have been reported in the literature.

In this study, we assume that the clathration efficiency is total, implying that guest molecules had the time to diffuse through porous water-ice solids before their growth into planetesimals and their accretion by proto-planets or proto-satellites. This statement remains plausible only if collisions between planetesimals have exposed essentially all the ice to the gas over time scales shorter or equal to planetesimals lifetimes in the nebula\cite{Lunine1985}. In this case, NH$_3$, H$_2$S, Xe, CH$_4$ and $\sim$ 38.6\% of CO form NH$_3$-H$_2$O hydrate and H$_2$S, Xe, CH$_4$ and CO clathrates with the available water in the outer nebula. The remaining CO, as well as N$_2$, Kr, and Ar, whose clathration normally occurs at lower temperatures, remain in the gas phase until the nebula cools enough to allow the formation of pure condensates. Note that, because we assume that the gas phase composition of the disk does not vary with the heliocentric distance, the calculated clathration conditions remain the same in the 5-20 AU range of the nebula, as shown recently by Marboeuf et al. \cite{Marboeuf08} in the case of their study of the composition of ices produced in protoplanetary disks. Once crystallized, these ices will agglomerate and form planetesimals large enough to decouple from the nebular gas and will be accreted by the forming planets and satellites.

\section{The statistical--thermodynamic model}
\label{model}

To calculate the relative abundances of guest species incorporated in a clathrate from a coexisting gas of specified composition at given temperature and pressure, we follow the method described by Lunine \& Stevenson\cite{Lunine1985} and Thomas et al.\cite{Thomas07,Thomas08,Thomas09} which uses classical statistical mechanics to relate the macroscopic thermodynamic properties of clathrates to the molecular structure and interaction energies. It is based on the original ideas of van der Waals \& Platteeuw for clathrate formation, which assume that trapping of guest molecules into cages corresponds to the three-dimensional generalization of ideal localized adsorption. This approach is based on four key assumptions\cite{Lunine1985,Sloan98}:

\begin{enumerate}
\item The host molecules contribution to the free energy is independent of the clathrate occupancy. This assumption implies in particular that the guest species do not distort the cages.

\item (a) The cages are singly occupied. (b) Guest molecules rotate freely within the cage.

\item Guest molecules do not interact with each other.

\item Classical statistics is valid, i.e., quantum effects are negligible.

\end{enumerate}

In this formalism, the fractional occupancy of a guest molecule $K$ for a given type $t$ ($t$~=~small or large) of cage can be written as

\begin{equation}
\label{occupation}
y_{K,t}=\frac{C_{K,t}P_K}{1+\sum_{J}C_{J,t}P_J} ,
\end{equation}

\noindent where the sum in the denominator includes all the species which are present in the initial gas phase. $C_{K,t}$ is the Langmuir constant of species $K$ in the cage of type $t$, and $P_K$ is the partial pressure of species $K$. This partial pressure is given by $P_K=x_K\times P$ (we assume that the sample behaves as an ideal gas), with $x_K$ the mole fraction of species $K$ in the initial gas phase given in Table \ref{lodders}, and $P$ the total gas pressure, which is dominated by H$_2$. 

The Langmuir constant depends on the strength of the interaction between each guest species and each type of cage, and can be determined by integrating the molecular potential within the cavity as

\begin{equation}
\label{langmuir}
C_{K,t}=\frac{4\pi}{k_B
T}\int_{0}^{R_c}\exp\Big(-\frac{w_{K,t}(r)}{k_B T}\Big)r^2dr ,
\end{equation}

\noindent where $R_c$ represents the radius of the cavity assumed to be spherical, $k_B$ the Boltzmann constant, and $w_{K,t}(r)$ is the spherically averaged Kihara potential representing the interactions between the guest molecules $K$ and the H$_2$O molecules forming the surrounding cage $t$. This potential $w(r)$ can be written for a spherical guest molecule, as\cite{McKoy63}

\begin{equation}
\label{pot_Kihara}
\begin{split} 
w(r) = 2z\epsilon\Big[\frac{\sigma^{12}}{R_c^{11}r}\Big(\delta^{10}(r)+\frac{a}{R_c}\delta^{11}(r)\Big) - \frac{\sigma^6}{R_c^5r}\Big(\delta^4(r)+\frac{a}{R_c}\delta^5(r)\Big)\Big],
\end{split}
\end{equation}

\noindent with

\begin{equation}
\delta^N(r)=\frac{1}{N}\Big[\Big(1-\frac{r}{R_c}-\frac{a}{R_c}\Big)^{-N}-\Big(1+\frac{r}{R_c}-\frac{a}{R_c}\Big)^{-N}\Big].
\end{equation}

\noindent In Eq. (\ref{pot_Kihara}), $z$ is the coordination number of the cell. This parameter, which depends on the structure of the clathrate (I or II) and on the type of the cage (small or large), is given in Table \ref{cages}. The Kihara parameters $a$, $\sigma$ and $\epsilon$ for the molecule-water interactions employed in this work have been taken from Diaz Pe\~na et al.\cite{DP82} for CO and from Parrish \& Prausnitz\cite{PP72a, PP72b} or from Sloan \& Koh\cite{Sloan08} for all the other molecules of interest (see Table \ref{kihara}).

\begin{table}
\caption[]{Parameters for the cavities\cite{PP72a, PP72b}.}
\begin{center}
\begin{tabular}{lcccc}
\hline
Clathrate structure & \multicolumn{2}{c}{I} & \multicolumn{2}{c}{II} \\
\hline
Cavity type     	& small     	& large     		& small     		& large \\
$R_c$ (\AA)     	& 3.975     & 4.300     	& 3.910     	& 4.730 \\
$b$             	& 2         	& 6         		& 16       		& 8     \\
$z$             	& 20        	& 24        		& 20        		& 28    \\
\hline
\end{tabular}
\label{cages}
\end{center}

$R_c$ is the radius of the cavity. $b$ represents the number of small ($b_s$) or large ($b_\ell$) cages per unit cell for a given structure of clathrate (I or II), $z$ is the coordination number in a cavity.
\label{cages}
\end{table}

Finally, the mole fraction $f_K$ of a guest molecule $K$ in a clathrate can be calculated with respect to the whole set of species considered in the system as

\begin{equation}
\label{abondance} f_K=\frac{b_s y_{K,s}+b_\ell y_{K,\ell}}{b_s \sum_J{y_{J,s}}+b_\ell \sum_J{y_{J,\ell}}},
\end{equation}

\noindent where $b_s$ and $b_l$ are the number of small and large cages per unit cell respectively, for the clathrate structure under consideration, and with $\underset{K}{\Huge{\Sigma}} f_{K}~=~1$.

\begin{table}
\caption[]{Two different sets for the Kihara potential.}
\begin{center}
\begin{tabular}{lccccc}
\hline
Set     	& Ref. 			& Molecule   	& $\sigma$ (\AA)	& $ \epsilon/k_B$ (K)	& $a$ (\AA) \\
\hline
(1)		& PP72			& H$_2$S		& 3.1558			& 205.85		& 0.36 	\\
		& PP72 + SK08		& CO$_2$	& 2.9681			& 169.09		& 0.6805  \\
	        	& PP72			& CH$_4$     	& 3.2398     		& 153.17     	& 0.300 	\\
        		& PP72			& N$_2$       	& 3.2199     		& 127.95     	& 0.350 	\\
        		& PP72			& Xe              	& 3.1906     		& 201.34     	& 0.280 	\\
        		& PP72			& Ar              	& 2.9434     		& 170.50     	& 0.184 	\\
        		& PP72			& Kr             	& 2.9739     		& 198.34     	& 0.230 	\\
       		& DP82			& CO            	& 3.101      		& 134.95     	& 0.284 	\\
		&				&			&				&			&	      	\\
(2)		& SK08			& H$_2$S		& 3.10000			& 212.047		& 0.3600 	\\
		& SK08			& CO$_2$	& 2.97638			& 175.405		& 0.6805 	\\
	        	& SK08			& CH$_4$     	& 3.14393     		& 155.593     	& 0.3834	 \\
        		& SK08			& N$_2$       	& 3.13512     		& 127.426     	& 0.3526	 \\
        		& SK08			& Xe              	& 3.32968     		& 193.708     	& 0.2357 	\\
        		& PP72			& Ar              	& 2.9434     		& 170.50     	& 0.184 	\\
        		& PP72			& Kr             	& 2.9739     		& 198.34     	& 0.230 	\\
       		& DP82			& CO            	& 3.101      		& 134.95     	& 0.284 	\\
\hline
\end{tabular}
\end{center}
$\sigma$ is the Lennard-Jones diameter, $\epsilon$ is the depth of the potential well, and $a$ is the radius of the impenetrable core. PP72, DP82 and SK08 correspond to the data taken from Parrish \& Prausnitz\cite{PP72a,PP72b}, Diaz Pe\~{n}a et al.\cite{DP82} and Sloan \& Koh\cite{Sloan08}, respectively. 
\label{kihara}
\end{table}

\section{Predicted clathrate hydrates occupancies in the primordial nebula}
\label{predict}

We calculate here the relative abundances of guests that can be incorporated in H$_2$S, Xe, CH$_4$ and CO clathrates at the time of their formation in the solar nebula. As far as we know, the two most complete sets of Kihara parameters available in the literature concerning astrophysical molecules are those published by Parrish \& Prausnitz and Diaz Pe\~na et al. The parameters given by Parrish \& Prausnitz have been obtained by comparing calculated chemical potentials based on the structural data of the clathrates cages with experimental results based on clathrate dissociation pressure data. These parameters have been used in recent work that aimed at investigating the composition of clathrates that may exist at the surfaces of Titan and Mars\cite{Thomas07,Thomas08,Thomas09,Swindle09}. On the other hand, the parameters given by Diaz Pe\~na et al. have been fitted to experimentally measured interaction virial coefficients for binary mixtures. These parameters have been used to quantify the trapping by clathrates of gases contained in volatiles observed in comets\cite{Iro03}. Thomas et al.\cite{Thomas08} did a comprehensive comparison between these two sets of data and concluded that the parameters given by Parrish \& Prausnitz are the most reliable because they have been self-consistently determined on experimentally measured clathrates properties and also because the results are similar to those obtained from other recent compilations (such as the one of Sloan \& Koh) whose molecules of astrophysical interest are listed in set (2) of Table \ref{kihara}). For these reasons, we have used the potential and structural parameters given by Parrish \& Prausnitz for H$_2$S, CH$_4$, N$_2$, Xe, Ar and Kr in our nominal calculations. However, in the case of CO, we have used the Kirara parameters from Diaz Pe\~na et al. because these data are absent from the compilation of Parrish \& Prausnitz. Finally, Thomas et al.\cite{Thomas09} found that, for common molecules, the parameters from Parrish and Prausnitz are very similar to those published in recent compilations also obtained from the fit of Langmuir constants to simple clathrate formation experimental data\cite{Sloan98,Jager2001,Kang01}, except for the value of the parameter $a$ for CO$_2$ (the radius of the impenetrable core), which is almost twice larger in the recent sets of parameters than in Parrish and Prausnitz's parameters. Following the conclusions of Thomas et al.\cite{Thomas09}, we have used the CO$_2$ Kirara parameters from Parrish \& Prausnitz, except for the suspicious $a_{CO_2}$ value, which has been replaced by the one given by Sloan \& Koh. Set (1) of Table \ref{kihara} enumerates the list of Kihara parameters for various molecules considered in our system and used in our nominal calculations.

\begin{figure}
\centering
\resizebox{\hsize}{!}{\includegraphics[angle=0]{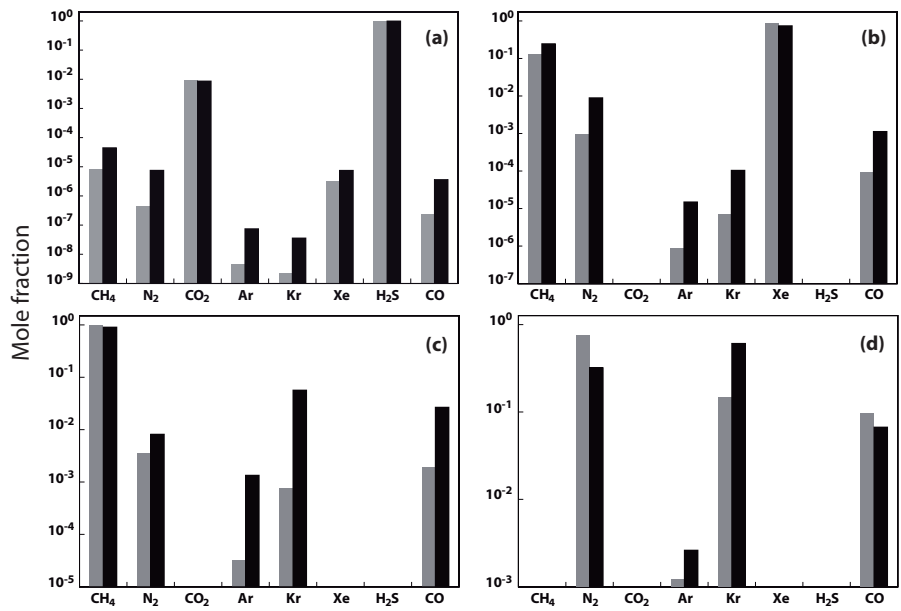}}
\caption{Mole fraction of volatiles encaged in H$_2$S (a), Xe (b), CH$_4$ (c) and CO (d) clathrates. Grey and dark bars correspond to structure I and structure II clathrates, respectively.}
\label{comp}
\end{figure}

In our calculations, any volatile already trapped or condensed at a higher temperature than the formation temperature of the clathrate under consideration is excluded from the coexisting gas phase composition. This implies that CO$_2$, Xe, CH$_4$, CO, Kr, Ar and N$_2$ are considered as possible guests in the case of H$_2$S clathrate. On the other hand, only N$_2$, Ar, Kr can become guests in CO clathrate. Figure \ref{comp} represents the mole fraction $f$ (Eq. 5) of volatiles encaged in structure I and structure II clathrates \textit{a priori} dominated by H$_2$S, Xe, CH$_4$ and CO and formed in the primordial nebula. Interestingly enough, this figure shows that,  contrary to H$_2$S, Xe and CH$_4$, which remain the dominating guest species in the clathrates considered, CO becomes a minor compound in the clathrate that is expected to be dominated by this molecule. Indeed, whatever the structure considered, Kr and N$_2$ become more abundant than CO in this clathrate, irrespective of their initial abundances in the gas phase of the nebula. This behavior results from the fact that the interaction potential between CO and the water molecules forming the surrounding cages is weaker than those involving the other species.
 
\begin{figure}
\centering
\resizebox{\hsize}{!}{\includegraphics[angle=0]{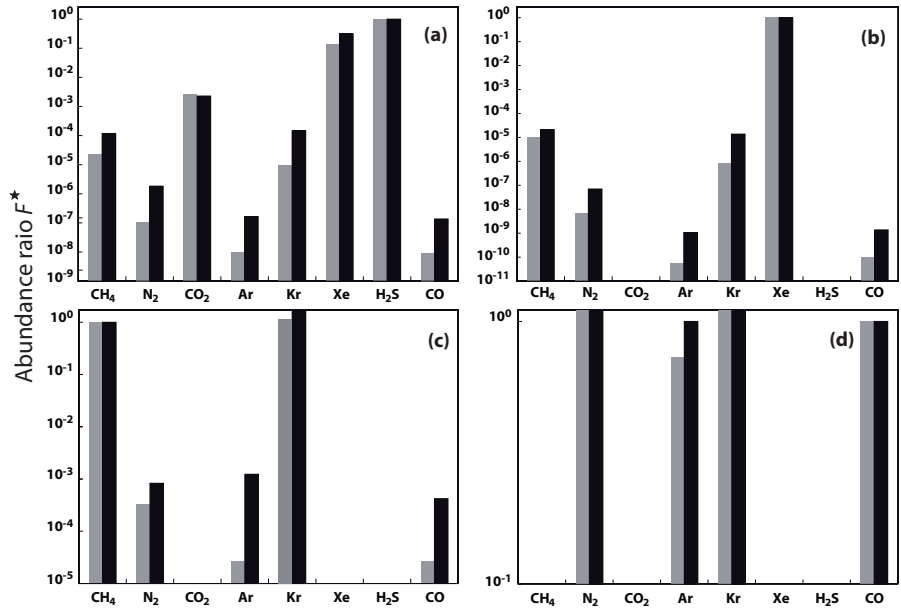}}
\caption{Abundance ratio $F^{\star}_{\rm K}$ is defined as the ratio of K/X in X clathrate (i.e. $f^{\star}_{\rm K}$) to K/X in the initial nebula gas phase (where X = H$_2$S (a), Xe (b), CH$_4$ (c) and CO (d)). Grey and dark bars correspond to structure I and structure II clathrates, respectively.}
\label{comp2}
\end{figure}

Table \ref{trapping} gives the temperature and pressure formation conditions of these clathrates in the nebula and the relative abundance $f^{\star}_{\rm K}$ defined as the ratio of K/X in X clathrate (where X = H$_2$S, Xe, CH$_4$ or CO). Figure \ref{comp2} represents the abundance ratio $F^{\star}_{\rm K}$ which is defined as the ratio of K/X in X clathrate (i.e. $f^{\star}_{\rm K}$) to K/X in the initial nebula gas phase. A guest K which incorporates completely into a given clathrate displays a $F^{\star}_{\rm K}$ value of 1 or greater, provided that there is enough water in the nebula available for clathration. This figure shows that volatile species such as Kr, Ar or N$_2$ that are expected to form pure condensates below 30 K in the primitive nebula can be efficiently encaged in clathrates formed at higher temperature ($\sim$ 48.5 K and above). Indeed, irrespective of the particular structure, and depending on the amount of water available for clathration, these three volatiles can be efficiently trapped in CO clathrate at $T$ = 48.5 K in the primitive nebula (panel (d) of Fig. \ref{comp2}).  Kr can also be efficiently enclathrated at at $T$ = 55.5 K in CH$_4$ clathrate (panel (c) of Fig. \ref{comp2}). Generally speaking, Figure \ref{comp2} illustrates the fact that some volatiles can be efficiently trapped at relatively high temperature in multiple guest clathrates compared to the temperatures at which they are expected to condense or to form single guest clathrates in the primordial nebula. It is important to note that the present results strongly depend on the choice of the initial gas phase conditions and on the amount of water ice available for clathration in the formation zone of planetesimals.

\begin{table}
\caption[]{Relative abundance $f^{\star}$ of volatiles encaged in structure I and structure II clathrates.}
\begin{center}
\begin{tabular}{lccc}
\hline
Clathrate					& Species			& Structure I			& Structure II		\\
\hline
 						&  			&   \multicolumn{2}{c}{$f^{\star}_{\rm K}$}	\\
\hline
H$_2$S			 		& CO$_2$		& $9.67 \times 10^{-3}$	& $8.74 \times 10^{-3}$	\\
$T$ = 82.3 K				& Xe				& $3.28 \times 10^{-6}$	& $7.58 \times 10^{-6}$	\\
$P$ = $1.7 \times 10^{-7}$ bar 	& CH$_4$      		& $8.46 \times 10^{-6}$     & $4.51 \times 10^{-5}$	\\
						& CO			& $2.47 \times 10^{-7}$	&  $3.66 \times 10^{-6}$	\\
						& Kr				& $2.35 \times 10^{-9}$	& $3.63 \times 10^{-8}$	\\
						& Ar				& $4.48 \times 10^{-9}$	& $7.54 \times 10^{-8}$	\\
						& N$_2$			& $4.50 \times 10^{-7}$	& $7.62 \times 10^{-6}$	\\
\hline
Xe						& CH$_4$      		& $1.52 \times 10^{-1}$	& $3.32 \times 10^{-1}$	\\
$T$ = 59.8 K				& CO			& $1.07 \times 10^{-4}$	& $1.52 \times 10^{-3}$	\\
$P$ = $1.2 \times 10^{-7}$ bar	& Kr				& $8.30 \times 10^{-6}$	& $1.41 \times 10^{-4}$	\\
						& Ar				& $1.01 \times 10^{-6}$	& $2.02 \times 10^{-5}$	\\
						& N$_2$			& $1.13 \times 10^{-3}$	& $1.20 \times 10^{-2}$	\\
\hline
CH$_4$					& CO			& $1.91 \times 10^{-3}$	& $2.95 \times 10^{-2}$	 \\
$T$ = 55.3 K				& Kr				& $7.64 \times 10^{-4}$	& $6.27 \times 10^{-2}$	\\
$P$ = $1.1 \times 10^{-7}$ bar	& Ar				& $3.24 \times 10^{-5}$	& $1.48 \times 10^{-3}$	\\
						& N$_2$			& $3.62 \times 10^{-3}$	& $9.01 \times 10^{-3}$	\\
\hline
CO						& Kr				& 1.51				& 9.08				\\
$T$ = 47.0 K				& Ar				& $1.25 \times 10^{-2}$	& $3.90 \times 10^{-2}$	\\
$P$ = $1.0 \times 10^{-7}$ bar	& N$_2$			& 7.84				& 4.78				\\
\hline
\end{tabular}
\end{center}
$f^{\star}_{\rm K}$ is defined as the relative abundance of guest K to X in X clathrate (where X = H$_2$S, Xe, CH$_4$ or CO). Values of $T$ and $P$ correspond to the temperature and pressure of the H$_2$-dominated gas at which the cooling curve (here at 5 AU) intercepts the equilibrium curves of the considered clathrates (see Fig. \ref{cool}) Only the species that are not yet condensed or trapped prior the epoch of clathrate formation are considered in our calculations. 
\label{trapping}
\end{table}

\section{Sensitivity to parameters}
\label{sens}

The determination of the relative abundances of guest species incorporated in clathrates formed in the primitive nebula, and more generally in clathrates formed in any thermodynamic condition, strongly depends on their structural characteristics (i.e. size of the cages) and also on the parameters of the Kihara potential. Here, we investigate the influence of these structural characteristics and potential parameters on the fractional occupancies of guests in clathrates formed in the nebula.

\subsection{Influence of the interaction potential parameters}

\begin{figure}
\centering
\resizebox{\hsize}{!}{\includegraphics[angle=0]{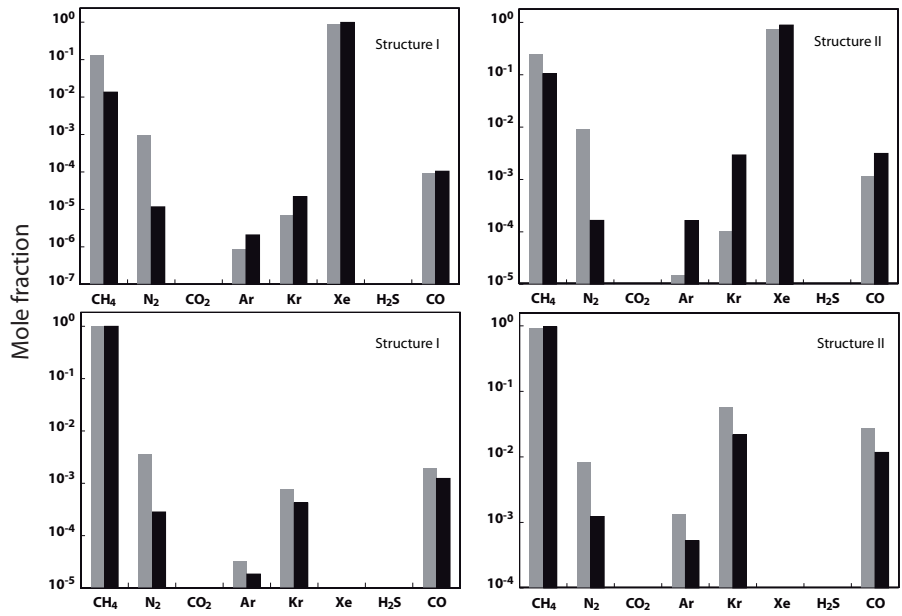}}
\caption{Mole fraction of volatiles encaged in structure I and structure II clathrates dominated by Xe (top panels) and CH$_4$ (bottom panels). Grey and dark bars correspond to sets (1) and (2) of parameters of the Kihara potential given in Table \ref{kihara}, respectively.}
\label{param}
\end{figure}

It has been recently shown that, perturbing in the 1--10\% range the $\sigma$ and $\epsilon$ Kihara parameters taken from a given compilation, leads to strong variations of the values of the Langmuir constants and thus of the fractional occupancies of enclathrated molecules\cite{papadimitriou07}. In the present case, in order to investigate the sensitivity of the composition of clathrates formed in the nebula to the variation of  Kihara parameters, we have tested a second compilation of parameters close to our nominal set of values taken from Parrish \& Prausnitz (set 1 of Table \ref{kihara}) but which derives from Sloan \& Koh for H$_2$S, CO$_2$, CH$_4$, N$_2$ and Xe. Because we did not find any Kihara parameter for Ar, Kr and CO in the recent published compilations, we have adopted the same values as in our nominal set for these molecules (see set 2 of Table \ref{kihara}). As a result, the two sets of of Kihara parameters used in our comparison are almost identical since several volatiles share the same data and the dispersion of data between other compounds is narrow. Two comparisons, represented in Figure \ref{param}, have been made in the cases of Xe and CH$_4$ clathrates produced in the primordial nebula. In both cases, calculations have been performed for structure I and II clathrates. Despite the similarity of the potential parameters used, one can note dispersions up to more than one order of magnitude in the resulting mole fractions (Eq. 5) for a given volatile encaged in the same clathrate. We then conclude that even minor dispersions between the different existing sets of Kihara parameters can lead to large variations into the determination of the volatiles trapping in clathrates formed in the primordial nebula. However, these variations were not found to be strong enough to modify the relative trapping efficiencies between the different volatiles in the clathrates considered here.

\subsection{Influence of cage variations}

In the present work, we have adopted the structural parameters of clathrates given by Parrish \& Prausnitz and shown in Table \ref{cages}. Up to now, we have assumed that the size of the cages $R_c$ is unaffected by the trapping conditions of the different guests in the primordial nebula. However, laboratory measurements have shown that the size of the cages could increase with temperature and also with the size of the incorporated guest species\cite{Shpakov97,Takeya06,Belosludov03,Hester07}.We have thus investigated the influence of variations of the cage sizes on the mole fractions of guests encaged in clathrates by modifying by up to $\pm$5\% the values of $R_c$ given in Table \ref{cages}. This large variation is consistent with typical thermal expansion or contraction measured in the temperature range 90--270 K \cite{Shpakov97,Takeya06,Belosludov03,Hester07}. Note that because clathration of volatiles occurs at lower temperature in the nebula, slightly  larger variations of the cage sizes may be expected.

\begin{figure}
\centering
\resizebox{\hsize}{!}{\includegraphics[angle=0]{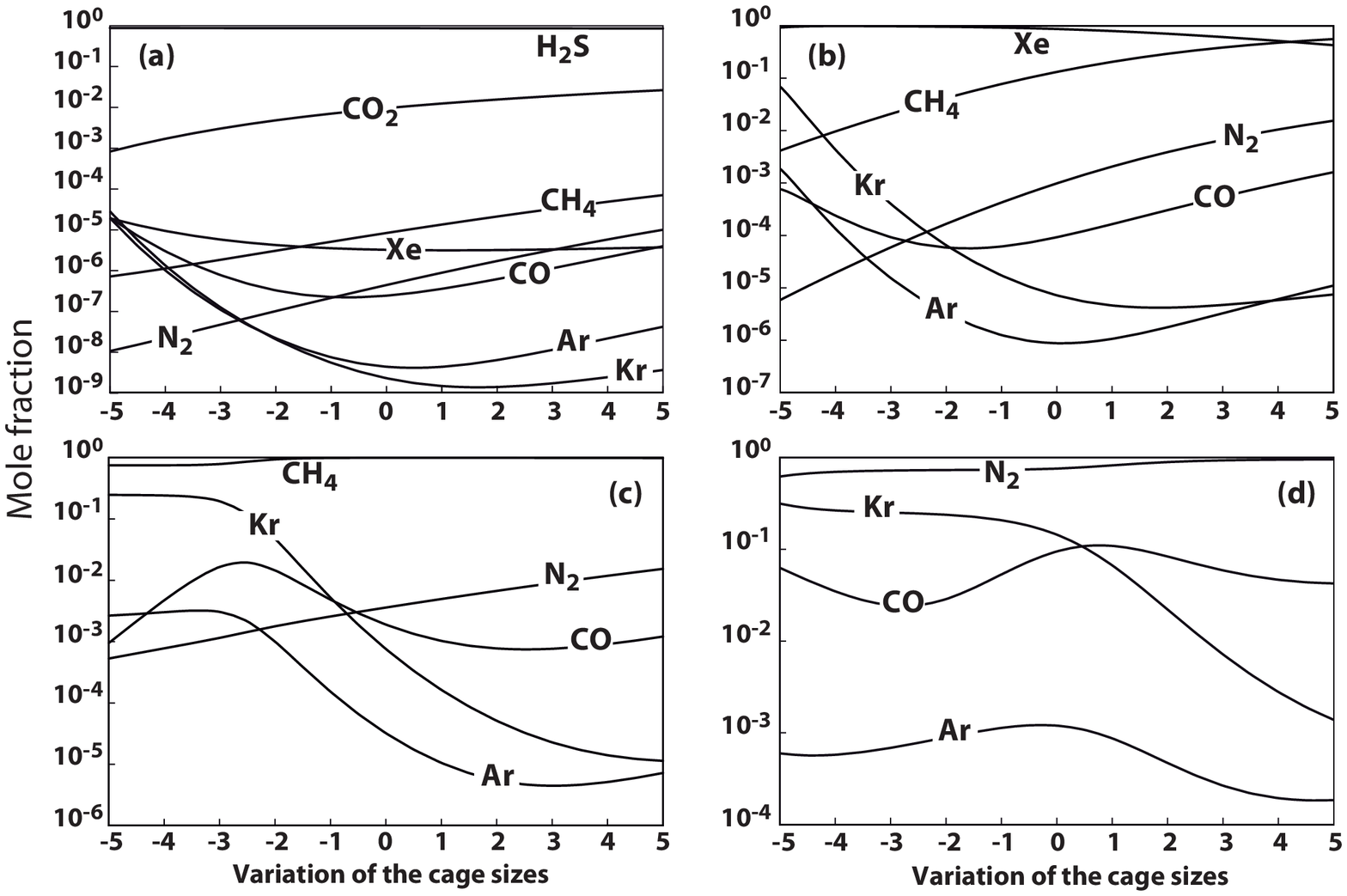}}
\caption{Mole fraction of volatiles encaged in structure I clathrates dominated by H$_2$S (a), Xe (b), CH$_4$ (c) and CO (d) as a function of the cage sizes.}
\label{var}
\end{figure}

The evolution of the mole fractions of all guests in structure I clathrates formed in the primitive nebula is given in Figure \ref{var} as a function of the size of the cages, at clathration temperatures and pressures given in Table \ref{trapping}. This figure shows that the contraction or expansion of the cages clearly affects the mole fractions of some volatiles in clathrates and that the magnitude of these changes strongly depends on the interaction parameters between the guest species and the cages. Indeed, irrespective of the clathrate considered, Figure \ref{var} shows that the mole fractions of Ar and Kr can vary up to several orders of magnitude by changing the size of the cages in the range considered. Moreover, depending on the particular clathrate, mole fractions of other molecules such as N$_2$ or CO are also subject to strong changes with the variation of the size of the cages. Similar trends have been revealed by performing calculations for clathrates of structure II. Our calculations confirm that the trapping propensity of a given molecule in clathrates is related rather sensitively to the shape of its intermolecular potential, which is determined from the adopted Kihara parameters and also from the adopted size of the cages (see Eq. \ref{pot_Kihara}). This is to be expected given that the Kihara parameters enter exponentially into the statistical partition function which determines the probability of occupancy. The tradeoff between the attractive part of the potential and the repulsive part is a function of the guest molecule shape and size, as well as the size of the cage, and given that clathrate structures feature two cage sizes, it is not easy to intuit what happens to the relative incorporation of a molecule as cages shrink or expand thanks to temperature. The statistical mechanical model is therefore essential for predictive purposes.

Figure \ref{var} suggests that thermal variations of the cages need to be taken into account in particular if these variations are greater than a few percent. Similar conclusions have been obtained by Thomas et al.\cite{Thomas08} in their study of clathrate formation and composition at the surface-atmosphere interface of Titan. As discussed in the following Section, these results may affect the predictions of the composition of the bodies formed in the outer solar system. On the other hand, variations with temperature are often not well constrained due to the small number of specific systems that have been studied, such as for example the pure methane clathrate for which the variations of the cages have been found to be small ($\sim$ 0.3 \%) between 83 and 173 K\cite{Takeya06}. In the present study, several species are encaged in the same clathrate and the temperature range of interest is lower than those considered in experiments. For these reasons, it is difficult to infer variation laws describing the expansion/contraction of clathrates at temperatures relevant for the solar nebula.

\section{Implications for the composition of the outer solar system}
\label{imp}
 
Our calculations of the multiple guest trapping in clathrates formed in the primordial nebula have implications for the formation and composition models of the giant planets, their surrounding satellites and also comets. Indeed, the fact that several compounds expected to be trapped or condensed at low temperature are incorporated at relatively higher temperature in clathrates in the nebula modifies the predictions of the composition of the icy planetesimals from which these bodies were presumably formed relative to what is obtained from simple condensation models. A few examples are discussed here.

\subsection{Volatile enrichments in Saturn}

Measurements by the mass spectrometer aboard the Galileo probe have shown that the abundances of C, N, S, Ar, Kr and Xe are all enriched by similar amounts with respect to their solar abundances in the atmosphere of Jupiter\cite{Owen99,Mahaffy00,Wong04}. Similarly, recent Cassini CIRS observations have also confirmed what was inferred from previous measurements, that C is substantially enriched in the atmosphere of Saturn\cite{Flasar05,Fletcher09}. In order to interpret these volatile enrichments, it has been proposed that the main volatile compounds initially existing in the solar nebula gas phase were essentially trapped by crystalline water ice in the form of clathrates or hydrates in the feeding zones of Jupiter and Saturn\cite{Gautier01a,Gautier01b,Alibert05,Hersant04,Hersant08,Mousis06,Mousis09b}. These ices then agglomerated and formed planetesimals that were ultimately accreted by the forming Jupiter and Saturn. This is then the fraction of these icy planetesimals that vaporized when entering the envelopes of the two growing planets which engendered the observed volatile enrichments.

On the other hand, our statistical model allows us to infer that multiple guest clathrates were more likely formed than single clathrates in the solar nebula, implying substantial changes in the presumed composition of the planetesimals formed from these ices. As a result, the volatile enrichments calculated in the giant planets atmospheres should also be altered because these quantities are proportional to the mass of accreted and vaporized icy planetesimals. Indeed, for instance, a recent interpretation of the carbon abundance in the atmosphere of Saturn is based on the hypothesis that the giant planet never formed at a disk temperature below 30 K\cite{Hersant08}, implying that the planetesimals accreted by the giant planet were impoverished in Ar, Kr, CO and N$_2$ (i.e. volatiles whose condensation curves are located below 30 K in the nebula -- see Fig. \ref{cool}). However, our nominal calculations predict that Kr can be entirely trapped in CH$_4$-dominated clathrates at $\sim$ 55 K in the nebula, provided that there is enough water available for clathration at this temperature in the feeding zone of Saturn. In this scenario, the trapping of Kr at higher temperature in the planetesimals accreted by Saturn implies that the atmospheric abundance of this noble gas should also be enhanced by an amount similar to that of carbon compared to solar (at least $\sim$ 9 times solar\cite{Fletcher09,Mousis09b}), instead of being predicted in solar abundance in Saturn's envelope\cite{Hersant08}. However, this conclusion is valid only for the structural characteristics and the Kihara parameters adopted in our nominal calculations. As shown in Figure \ref{var}, any expansion of the cage sizes during the formation of CH$_4$-dominated clathrate in the solar nebula could strongly reduce the trapping efficiency of Kr in this clathrate and thus decrease its resulting enhancement in the atmosphere of Saturn.

\subsection{The argon deficiency in Titan}

A puzzling feature of the atmosphere of Titan is that no primordial noble gases other than argon were detected by the Gas Chromatograph Mass Spectrometer (GCMS) aboard the Huygens probe during its descent to Titan's surface on January 14, 2005. The observed argon includes primordial ${}^{36}$Ar, i.e. the main isotope, and the radiogenic isotope ${}^{40}$Ar, which is a decay product of ${}^{40}$K\cite{Niemann05}.  In any case, the ${}^{36}$Ar/${}^{14}$N is lower than the solar value by more than five orders of magnitude\cite{Niemann05}. The other primordial noble gases Kr and Xe (and  $^{38}$Ar) were not detected by the GCMS instrument down to upper limits of 10 parts per billion relative to nitrogen\cite{Niemann05}. The Kr and Xe deficiencies could simply be explained by the presence of clathrates on the surface of Titan that would have efficiently incorporated these noble gases\cite{Thomas07,Thomas08}. On the other hand, in order to interpret the Ar deficiency in the atmosphere of Titan, it has been proposed that the satellite was formed from icy planetesimals initially produced in the solar nebula and that were partially devolatilized at a temperature not exceeding $\sim$ 50 K during their migration within Saturn's subnebula\cite{Mousis09a}. In this case, because Ar is poorly trapped in clathrates formed above $\sim$ 50 K in the nebula, only tiny amounts of this compound would have been incorporated in the building blocks of the forming Titan, in agreement with the observations. In particular, our nominal model predicts that this noble gas would remain essentially trapped in CH$_4$-dominated clathrate\footnote{The amount of Ar found in Xe and H$_2$S-dominated clathrates is negligible.}, and subsequently in the satellite, with Ar/CH$_4$ of 2.7 $\times$ 10$^{-5}$ (see Fig. \ref{comp2}), in good match with the abundances of ${}^{36}$Ar observed in Titan's atmosphere. However, this statement remains valid only if one considers our nominal calculations of multiple guest trapping in clathrates formed in the feeding zone of Saturn. Indeed, Figure \ref{var} shows that the trapping of Ar can strongly increase by a factor of more than 300 in CH$_4$-dominated clathrate formed in the solar nebula if the size of the cages decreases by a few percent. In this case, the amount of argon predicted by our calculations in the atmosphere of Titan would be higher than the value measured by Huygens or an alternative scenario must be invoked to explain its apparent depletion.

\subsection{Noble gas content in comets}

It has been proposed that the composition of volatiles observed in comets could be explained on the basis of their trapping in the form of clathrates in the primordial nebula\cite{Iro03}. In this model, the key parameter to explain the volatile content observed in comets is the amount of water ice available in the region of the nebula where the clathration took place. The mass of available water is then varied between the amount needed to trap the whole mass of volatiles present in the nebula (high-mass ice scenario) and that which is just required to enclathrate volatiles at temperatures above $\sim$ 50 K in the nebula (low-mass ice scenario). In this model, the case of argon is interesting because its relative abundance in comets is found to be very small (Ar/H$_2$O $\sim$ 10$^{-8}$) when the low-mass ice scenario is considered. On the other hand, if enough water ice is present, Ar/H$_2$O jumps to potentially detectable values (Ar/H$_2$O $\sim$ 10$^{-4}$--10$^{-3}$). However, as in the case of Titan described above, taking into account the uncertainties of our calculations (in particular on the cage sizes), substantial amounts of Ar can be trapped in CH$_4$-dominated clathrates. In this case, we find Ar/H$_2$O $\sim$ 10$^{-5}$--10$^{-4}$ in the low-mass ice scenario, a value which is close to the one determined in the case of the high-mass ice scenario. Our estimate for the low-mass ice scenario implies then that it is difficult to consider the measurement of the abundance of argon in comets as a key test to constrain the mass of water ice that was available in the nebula for forming comets.

\section{Summary and discussion}
\label{sum}

In this paper, we have calculated the relative abundances of guest species that can be incorporated in clathrates formed in the gas phase and thermodynamic conditions of the primordial nebula. We have assumed that the clathration efficiency is total in the primitive nebula, implying that guest molecules had the time to diffuse through porous water-ice solids before their growth into planetesimals and their accretion by proto-planets or proto-satellites. This statement remains plausible only if collisions between planetesimals have exposed essentially all the ice to the gas over time scales shorter or equal to planetesimals lifetimes in the nebula\cite{Lunine1985}. However, it is important to note that the efficiency of collisions between planetesimals to expose all the ``fresh'' ice over such a time scale still remains questionable and that we have no evidence that clathration was important in the primordial nebula.

The results presented here derive from the usual statistical model based on the van der Waals and Platteeuw approach, generalized by Parrish \& Prausnitz for the calculations of dissociation pressures of multiple guest clathrates. The major ingredient of our model is the description of the guest-clathrate interaction by a spherically averaged Kihara potential with a nominal set of parameters, most of which being fitted on experimental equilibrium data. Our model allows us to find that Kr, Ar and N$_2$ can be efficiently encaged in clathrates formed at temperatures higher than $\sim$ 48.5 K, the temperature of CO clathrate, in the primitive nebula, instead of forming pure condensates below 30 K. However, we find at the same time that the determination of the relative abundances of guest species incorporated in these clathrates strongly depends on the choice of the parameters of the Kihara potential and also on their adopted structural characteristics. Indeed, testing different potential parameters, we have noted that even minor dispersions between the different existing sets can lead to non negligible variations in the determination of the volatiles trapped in clathrates formed in the primordial nebula. However, these variations are not found to be strong enough to reverse the relative abundances between the different volatiles in the considered clathrates. Moreover, we have found that the contraction or expansion of the cages, due to temperature variations, can alter the mole fractions of some volatile molecules up to  several orders of magnitude in clathrates. On the other hand, due to the lack of laboratory experiments describing properly the variation of the size of cages as a function of temperature for molecules of astrophysical interest, it is actually difficult to quantify the influence of this parameter on the composition of planetesimals formed in the nebula. The determination of specific laws describing the variation with temperature of the size of cages is important because it could help to constrain the thermodynamic conditions encountered by planetesimals during their formation in the primordial nebula. Indeed, given a sufficient number of species whose abundances are determined in a particular object, our statistical mechanical model could be used to predict the composition and temperature (through the thermal expansion/contraction of the ice which affects cage size) of the planetesimals from which the object formed and hence the composition of bodies issued from the same family. Such a measurement will be achievable by the Rosetta spacecraft mission towards Comet 67P/Churyumov-Gerasimenko or by the next generation of missions designed to investigate the outer solar system.

Our results alter the predictions of the composition of the planetesimals formed in the outer solar system. In particular, the volatile abundances calculated in the giant planets atmospheres should be recalculated because these quantities are proportional to the mass of accreted and vaporized icy planetesimals. For the same reasons, the estimates of the volatile budgets accreted by icy satellites and comets must be reconsidered based on our calculations. For instance, our calculations predict that the abundance of argon in the atmosphere of Titan might for some initial conditions be higher than the value measured by Huygens because substantial amounts of this volatile may be trapped in the satellite's building blocks during their formation. Similarly to the case of Titan, if comets were agglomerated from clathrates in the nebula, then the Ar abundance in these bodies should still remain potentially detectable, irrespective of their formation temperature ($\sim$ 50 K and below) because substantial amounts of this volatile are expected to be trapped in CH$_4$-dominated clathrate at $\sim$ 55.5 K.

It is important to note that the validity of the statistical model to determine the composition of clathrates in the nebula probably meets some limitations. In particular, the transferability of the Kihara parameters to temperatures and pressures beyond the range for which they have been fitted is uncertain\cite{Ballard00,Clarke03}. Moreover, inconsistencies have been evidenced between Kihara parameters derived from different sets of experimental data\cite{Tee66}. In these conditions, supplementary measurements  of the equilibrium pressure of clathrates at low temperature would be needed in order to obtain a set of Kihara parameters consistent with the low temperature and pressure conditions of the nebula. Moreover, it has been claimed in the literature that Kihara potential may not accurately describe the interaction between guest and water molecules\cite{Sun05}. Recent works have thus been based on an atom-atom description of the intermolecular guest-clathrate interactions, in which effective parameters for these interactions have been fitted from results of ab initio quantum mechanical methods\cite{Sun05,Klauda02,Klauda03}. However, the accuracy of this atom-atom approach is strongly dependent on the ingredients of the fit, which can be the number of water molecules considered in the ab initio calculations, the number of sites chosen to represent the interacting molecules, the level of accuracy of the quantum methods used... It is then unfortunately very difficult to discriminate which approach (atom-atom or spherically averaged) is the most accurate.

\bibliography{rsc}

\end{document}